\documentclass[twocolumn]{article}

\usepackage{amssymb}
\usepackage{graphicx}
\usepackage{amsmath}
\usepackage{xcolor}
\usepackage{natbib}
\usepackage{lscape}
\usepackage{mathtools}
\usepackage{hyperref}
\usepackage{lineno}
\usepackage[varg]{txfonts}

\usepackage{authblk}

\title{Dynamo Action of Jupiter's Zonal Winds}

\author[1]{Johannes Wicht} 
\author[2]{Thomas Gastine}
\author[3]{Lucia D.~V.~Duarte}
\author[1]{Wieland Dietrich}

\affil[1]{Max Planck Institute for Solar System Research, 
Justus-von-Liebig-Weg 3, 37077 G\"ottingen, Germany}
\affil[2]{College of Engineering, Mathematics and Physical Sciences,  University of Exeter, 
Physics building,
Stocker Road,
Exeter, EX4 4QL, 
United Kingdom}
\affil[3]{IPGP, 
Institution for Higher Education and Research 1, rue Jussieu, 
75238 Paris cedex 05, France}

\renewcommand{\widetilde}{\tilde}
\renewcommand{\widehat}{\hat}

\newcommand{\dint}{\mathrm{d}}

\newcommand{\dissi}{\varphi}
\newcommand{\Dissi}{\varPhi}

\newcommand{\Ent}{\varTheta}

\newcommand{\refp}[1]{(\ref{#1})}

\newcommand{\Bv}{\mbox{$\boldsymbol{B}$}}
\newcommand{\Uv}{\mbox{$\boldsymbol{U}$}}
\newcommand{\Ev}{\mbox{$\mathbf E$}}

\newcommand{\jv}{\mbox{$\boldsymbol{j}$}}

\newcommand{\U}{\mbox{U}}
\newcommand{\B}{\mbox{B}}

\newcommand{\Uz}{\mbox{$\overline{U}_\phi$}}

\newcommand{\be}{\begin{equation}}
\newcommand{\ee}{\end{equation}}
\newcommand{\eep}{\;\;.\end{equation}}
\newcommand{\eec}{\;\;,\end{equation}}
\newcommand{\bea}{\begin{eqnarray}}
\newcommand{\eea}{\end{eqnarray}}
\newcommand{\bel}[1]{\begin{equation}\label{#1}}
\newcommand{\beal}[1]{\begin{eqnarray}\label{#1}}
\newcommand{\tp}[1]{\mbox{$\times10^{#1}$}}
\newcommand{\curl}{{\boldsymbol{\nabla}}\times}
\newcommand{\uvr}{\hat{\mathbf r}}
\newcommand{\uvp}{\hat{\mathbf \phi}}
\newcommand{\uvt}{\hat{\mathbf \theta}}

\newcommand{\Rm}{\mbox{Rm}}

\newcommand{\RmL}{\mbox{Rm$^{(1)}$}}
\newcommand{\RmLS}{\mbox{Rm$^{(2)}$}}
\newcommand{\SDCR}{SDCR}

\newcommand{\rhob}{\widetilde{\rho}}
\newcommand{\Tb}{\widetilde{T}}

\newcommand{\Figref}[1]{Fig.~\ref{#1}}
\newcommand{\figref}[1]{fig.~\ref{#1}}

\newcommand{\Eqnref}[1]{Eqn.~(\ref{#1})}
\newcommand{\eqnref}[1]{eqn.~(\ref{#1})}

\newcommand{\Secref}[1]{Sect. \ref{#1}}
\newcommand{\secref}[1]{sect. \ref{#1}}

\newcommand{\tabref}[1]{tab. \ref{#1}}
\newcommand{\Tabref}[1]{Tab. \ref{#1}}

\newcommand{\revi}[1]{#1}

\begin{document}

\maketitle

\begin{abstract}
{The new data delivered by NASA's Juno spacecraft significantly increase our understanding of Jupiter's internal dynamics.  
The gravity data constrain the depth of the zonal flows observed at cloud level and suggest that they slow 
down considerably at a depth of  about $0.96\,r_J$, 
where $r_J$ is the mean radius at the one bar level. 
Juno's magnetometer reveals the planet's internal magnetic  field.}  
{We combine the new zonal flow and magnetic field models with an updated electrical conductivity profile to assess the 
zonal wind induced dynamo action, concentrating on the outer part of Jupiter's molecular hydrogen region where the conductivity increases
very rapidly with depth.}
{Dynamo action remains quasi-stationary and can thus reasonably be 
estimated where the magnetic Reynolds number remains 
smaller than one, which is roughly the region above 
$0.96\,r_J$.
We calculate that the locally induced radial magnetic field reaches rms 
values of about $10^{-6}\,$T in this region 
and may just be detectable by the Juno mission. 
Very localized dynamo action and a distinct pattern that 
reflects the zonal wind system increases the chance to disentangle this 
locally induced field from the background field.
The estimates of the locally induced currents also allow calculating  
the zonal flow related Ohmic heating and associated entropy production. 
\revi{The respective quantities remain below new revised predictions 
for the total dissipative heating and total entropy production in 
Jupiter for any of the explored model combinations. 
Thus neither Ohmic heating nor entropy production offer additional
constraints on the depth of the zonal winds.}}
\end{abstract}

\section{Introduction}
\label{Introduction}

Two of the main objectives of NASA's Juno mission are to 
measure Jupiter's magnetic field with unprecedented resolution 
and to determine the depth of the fierce zonal winds observed 
in the planet's cloud layer. 
The first Juno-based internal magnetic field model JRM09 
\citep{Connerney2018} already provides the internal magnetic field up to
spherical harmonic degree $10$ and shows several interesting 
features that seem unique to Jupiter's dynamo \citep{Moore2018}. 
Better resolved models are expected as the mission continues. 

Based on Juno gravity measurements \citep{Iess2018}, 
\citet{Kaspi2018} deduce that the speed of the 
equatorially antisymmetric zonal flow contributions must 
be significantly reduced at a depth of about $3000\,$km
below the one bar level, which corresponds to a radius if $0.96\,r_J$. 
\citet{Kong2018} come to roughly similar conclusions  
with a different inversion procedure, but they also point out 
that the solution is not unique. 
While the gravity data only allow constraining the 
equatorially antisymmetric winds, the results likely 
also extend to the symmetric contributions. 
New interior models \citep{Guillot2018,Debras2019} 
and also the width of the dominant equatorial jet 
\citep{Gastine2014,Heimpel2016} both support 
the idea that the fast zonal winds are roughly
confined \revi{to the outer} $4$\% in radius.  

The fast planetary rotation enforces geostrophic   
flow structures with minimal variation 
along the direction of the rotation axis. Geostrophic zonal  
winds are thus expected to reach right through the planet's gaseous 
envelope, and it remains unclear which mechanism 
limits their extend in Jupiter. 
The demixing of hydrogen and helium and the subsequent precipitation 
of helium deeper into the planet offers 
one possible explanation \citep{Militzer2016}. 
This process would have established a helium gradient
that suppresses convection. In Jupiter, this stable helium-rain layer may 
start somewhere between $0.93$ and $0.90\,r_J$ and 
perhaps extends down to $0.80\,r_J$ \citep{Debras2019}. 
Note, however, that ab initio simulations by \citet{Schoettler2018} 
predict that the hydrogene/helium demixing may not even have started. 
Recent analysis of gravity measurements by the Cassini 
spacecraft suggest that Saturn's zonal winds 
may only reach down to about $0.85\,r_S$ \citep{Iess2019,Galanti2019}. 
Since the stably stratified layer is thought to start
significantly deeper, at about $0.62\,r_S$ according to 
\citep{Schoettler2018}, it cannot be the reason 
for this limited depth extend of Saturn's zonal winds. 

A second possibility to brake the zonal winds at depth are Lorentz forces.
Lorentz forces are tied to dynamo action and thus to
the electrical conductivity profile.  
Ab initio simulations for Jupiter suggest that ionization 
effects lead to a super-exponentially increase of the electrical 
conductivity in the outermost molecular gas envelope. 
We will refer to this layer as Jupiter's Steeply Decaying Conductivity 
Region (\SDCR) in the following. 
At about $0.9\,r_J$, hydrogen, 
the planet's main constituent, becomes metallic, and 
the conductivity increases much more smoothly with depth 
\citep{French2012} (see panel a) of \figref{fig:sigma}). 
Though dynamo action and the potential braking of the zonal winds 
due to Lorentz forces are classically attributed to the metallic 
region, they may already become significant where the 
electrical conductivity reaches sizable levels in the \SDCR.

Different dynamo-related arguments have been evoked to 
estimate the depth of the zonal winds without, however, 
directly addressing the role of the Lorentz forces. 
\citet{Liu2008} estimate
that the Ohmic heating caused by zonal-wind related induction  
would exceed the total heat emitted from Jupiter's interior, 
should the winds reach deeper than  $0.96\,r_J$ with undiminished 
speed. 
\citet{Ridley2016} argue that the secular variation 
of the magnetic field over 30 years of pre-Juno
observations is rather small and thus likely 
incompatible with an advection by undiminished zonal winds. 
They conclude that the winds cannot reach to depths 
where the magnetic Reynolds number exceeds one and 
more significant induction can be expected. This  
puts the maximum depth somewhere between $0.96\,r_J$ 
and $0.97\,r_J$, as we will discuss below.
\revi{A recent analysis by \citet{Moore2019} suggests that 
the observations over a $45\,$year time span 
including Juno data would by compatible with zonal wind 
velocities of $2.4\,$m/s at $0.95\,r_J$, 
two orders of magnitude smaller than observed in the 
cloud layer.}

Another interesting question is how much the dynamo action in the 
\SDCR\ contributes to Jupiter's total magnetic field.
Using a simplified mean-field approach, 
\citet{Cao2017} predict that the radial component 
of the Locally Induced 
Field (LIF) may reach $1\,$\% of the background field and 
could thus be detectable by the Juno magnetometer. 
\citet{Wicht2019} analyze the dynamo action in the
\SDCR\ of fully self-consistent numerical simulations that yield Jupiter-like 
magnetic fields. Because of the dominance of Ohmic diffusion, 
the dynamo dynamics becomes quasi-stationary 
in the \SDCR\ of their simulations.   
A consequence is that the locally induced electric 
currents and field can be estimated with decent precision 
when flow, electrical conductivity profile, and  
the surface magnetic field are known. 
Refined information on all three ingredients has recently become
available for Jupiter, allowing for a fresh look on the problem.  

Here we use three different zonal flow models, two electrical 
conductivity models, and the new Juno-based magnetic field model
JRM09 to predict the electric currents 
and magnetic fields produced in Jupiter's \SDCR. 
In addition, we also derive new estimates for the total dissipative 
heating and related entropy production and 
explore whether either value is exceeded by the zonal-flow
related Ohmic dissipation. 

The article starts off with outlining the methods and introducing 
the used data in 
\secref{sec:Methods}. \Secref{sec:Heat} discusses dissipative 
heating and entropy production in Jupiter. Estimates for dynamo 
action, Ohmic heating, and entropy production are then presented in \secref{sec:Heating}. 
\Secref{sec:Conclusion} closes the article with 
a discussion and conclusion. 
\section{Methods and Data}
\label{sec:Methods}

\subsection{Estimating Dynamo Action}
\label{sec:estimates}

The ratio of inductive to diffusive effects in the induction 
equation, 
\bel{eq:induct}
\frac{\partial \Bv}{\partial t} = \curl\left(\Uv\times\Bv\right) -
\curl \lambda \curl \Bv
\eec
can be quantified by the magnetic Reynolds number 
\bel{eq:Rm}
 \Rm = \frac{\langle\Uv\rangle\;D}{\lambda}
\eec
where $\lambda=1/(\mu\sigma)$ is 
the magnetic diffusivity, with $\mu$ the magnetic permeability 
and $\sigma$ the electrical conductivity. Angular brackets 
generally denote rms values at a given radius throughout the paper; 
thus $\langle\Uv\rangle$ stands for 
\bel{eq:rms}
\langle\Uv\rangle = \left( \frac{1}{4 \pi} \int_0^{2\pi}\dint \phi\;\int_0^\pi 
\dint \theta\;\sin\theta
\;\Uv^2\right)^{1/2}
\eec
$\theta$ being the colatitude and $\phi$ the longitude. 

The typical length scale $D$ is hard to estimate, and the  
planetary radius is often used for simplicity. 
Where $\sigma$ decreases steeply in 
the \SDCR, however, the length scale is determined by 
the conductivity or magnetic diffusivity scale height 
\bel{eq:dlambda}
D_\lambda=\frac{\lambda}{\partial\lambda / \partial r} = 
- \frac{\sigma}{\partial\sigma /\partial r}
\eec
and the modified magnetic Reynolds number 
\bel{eq:Rm1}
 \RmL = \frac{\langle\Uv\rangle\;D_\lambda}{\lambda}
\ee
should be used. Since $D_\lambda$ is small and 
$\lambda$ decreases steeply with radius, most of 
the \SDCR\ is characterized by a small magnetic Reynolds number
$\RmL<1$, and the magnetic field dynamics becomes quasi-stationary 
\citep{Liu2008}, \revi{obeying the simplified induction equation} 
\bel{eq:GS}
\curl \frac{\jv}{\sigma} \approx 
\curl\left(\Uv\times\widetilde{\Bv}\right)
\eep
Here, $\jv$ is the current density 
and $\tilde{\Bv}$ the strong background field produced by the 
dynamo acting deeper in the planet. 
The locally induced field $\widehat{\Bv}$ is given by 
Ampere's law:
\bel{eq:AL}
 \jv = \curl\widehat{\Bv}\,\big/\,\mu
\eep
The steep $\sigma$ profile dominates the radial dependence of 
$\jv$ and $\widehat{\Bv}$ in the \SDCR. 
The current density is thus dominated by the horizontal
components, where radial gradients in  
$\hat{\Bv}$ contribute \citep{Liu2008,Wicht2018}:
\bel{eq:JA}
\jv \approx \jv_H \approx \uvr\times\frac{\partial}{\partial r}\,\hat{\Bv}_H
\eep
Index $H$ denotes the horizontal components;  
the radial current  can be neglected in comparison. 

Along the same lines, \revi{the horizontal components of} \eqnref{eq:GS} can be approximated by 
\bel{eq:GSL}
\frac{1}{r}\frac{\partial}{\partial r} \frac{r}{\sigma}\;\jv_H 
\approx  -\uvr\times\left[\curl\left(\Uv\times\tilde{\Bv}\right)\right]_H
\eec
where $\uvr$ is the radial unit vector. 
Integration in radius yields the integral current density estimate 
\revi{introduced by \citet{Liu2008}}, which we 
identify with an upper index $(I)$: 
\bel{eq:JIH}
\jv^{(I)}_H = \frac{\sigma}{r} 
\left[\frac{r}{\sigma}\;\jv_H\right]_{r_J} + \uvr\times 
\frac{\sigma}{r}\;\int_r^{r_J}\,\dint r^\prime \;r^\prime\;
\left[\curl\left(\Uv\times\tilde{\Bv}\right)\right]_H
\eep
The square brackets with a lower index $r_J$ indicate 
that the expression
should be evaluated at the outer boundary. 

\revi{For a predominantly zonal flow, we can use the approximation} 
\beal{eq:CUxB}
\left[\curl\left(\Uv\times\tilde{\Bv}\right)\right]_H &\approx&
-\frac{\Uz}{r\sin\theta}\left(\frac{\partial}{\partial\phi}\,\tilde{B}_\theta\right)\,\uvt\\ \nonumber
&&+\frac{1}{r}\left[\frac{\partial}{\partial r}\,
\left(r\Uz\tilde{B}_r\right)
+\frac{\partial}{\partial\theta}\left(\Uz\tilde{B}_\theta\right)
\right]\,\uvp\;\;.
\eea
\revi{where $\Uz$ is the zonal flow component and 
$\uvt$ and $\uvp$ are unit vectors in latitudinal and azimuthal
direction, respectively.}

\revi{The integral estimates for the two horizontal current components
are then given by} 
\beal{eq:JIt}
j^{(I)}_\theta & = & \frac{\sigma}{r}\left[\frac{r}{\sigma}\;
j_\theta\right]_{r_J} - 
\frac{\sigma}{r}\;\left(\left[r\,\Uz \tilde{B}_r \right]_{r_J} -
                        \left[r\,\Uz \tilde{B}_r \right]_r \right)\\
&&\nonumber 
-\;\frac{\sigma}{r}\int_r^{r_J}\;\dint r^\prime\;
\frac{\partial}{\partial\theta}\,\left( \Uz\tilde{B}_\theta \right)    
\eea
and
\bel{eq:JIp}
j_\phi^{(I)} = \frac{\sigma}{r}\left[\frac{r}{\sigma}\;
j_\phi\right]_{r_J} 
-\frac{\sigma}{r}\;\int_r^{r_J}\;\dint r^\prime\;
\frac{\Uz}{\sin\theta}
\frac{\partial}{\partial\phi}\,\tilde{B}_\theta
\eep
Since the latitudinal length scale of the zonal winds is smaller
than the azimuthal length scale of the magnetic field, we 
expect that the latitudinal component dominates.  

The integral estimate requires the knowledge of the surface currents. 
While the surface currents are certainly very small,  
the scaled version $\sigma(r)/\sigma(r_J)\,\jv$ 
may remain significant. 
\citet{Liu2008} argue that neglecting the surface contribution
at least provides a lower bound for the rms current density. 

\citet{Wicht2019} confirm that the dynamics 
indeed becomes quasi-stationary in the \SDCR\ of Jupiter-like
dynamo simulations where $\RmL<1$ and show that 
$j_\theta$ is indeed the dominant current component 
in the \SDCR\ of their Jupiter-like dynamo simulations. 
They also report that the simplified Ohm's law 
for a fast moving conductor, 
\bel{eq:OLS}
\jv^{(O)} = \sigma \left(\Uv\times\tilde{\Bv}\right)
\eec
provides a significantly better estimate  
than $\jv^{(I)}$. 
We identify the respective current estimate with 
an upper index $(O)$.
\revi{The general Ohm law,} 
\bel{eq:OL}
\jv = \sigma \left(\Uv\times\Bv\,+\, \Ev\right)
\eec
\revi{
also contains currents driven by the electric field, which  
reduces to $\Ev=-\nabla \Phi$ in the quasi-stationary case, 
where $\Phi$ is the electric potential. 
In the \SDCR, this contribution likely proves secondary because 
the potential differences remain small compared to the 
induction by fast zonal winds \citep{Wicht2019}.} 

As the electrical conductivity decreases in the \SDCR,  
the magnetic field approaches a potential field with 
its characteristic radial dependence. 
We use this dependence to approximate the background 
field with 
\bel{eq:Bpot}
 \widetilde{\Bv}_\ell(r,\theta,\phi) \approx 
 \left( \frac{r_J}{r} \right)^{\ell+2}\;\Bv_\ell(r_J,\theta,\phi)
\eec
where the index $\ell$ denotes the magnetic field 
contribution at spherical harmonic degree $\ell$. 
This provides a decent approximation as long 
as the LIF remains a small contribution of the 
total field \citep{Wicht2019}.

Given a surface field model and an electrical conductivity 
profile, Ohm's law for a fast moving conductor and
a predominantly zonal flow suggests 
\bel{eq:OhmZ}
\jv\approx j^{(O)}_\theta = \sigma\;\Uz\;\tilde{B}_r
\eep
When using this result to constrain the outer-boundary currents, 
the alternative integral estimates, 
\eqnref{eq:JIt} and \eqnref{eq:JIp}, yield 
\bel{eq:JItO}
j^{(I)}_\theta = \sigma\;\Uz\;\tilde{B}_r \;-\;
\frac{\sigma}{r}\;\int_r^{r_J}\,\dint r^\prime\;\frac{\partial}{\partial\theta}\,
\left(\Uz\tilde{B}_\theta\right)
\ee
and
\bel{eq:JIpO}
j^{(I)}_\phi = 
-\frac{\sigma}{r}\;\int_r^{r_J}\,\dint r^\prime\;\frac{\Uz}{\sin\theta}
\frac{\partial}{\partial\phi}\,\tilde{B}_\theta
\eec
respectively.
A comparison of the estimates shows that $\jv^{(I)}$ and  $\jv^{(O)}$ 
will remain very similar at shallow depths. 
When the flow decays very deeply with depth, however, the integral contributions in 
\eqnref{eq:JItO} and \eqnref{eq:JIpO} will  
dominate below some radius and cause larger deviations,  
as we will see below.  

Calculating the LIF requires to uncurl Ampere's law, which  
reduces to integrating \eqnref{eq:JA} in the \SDCR. 
When using $\jv^{(O)}$, this yields    
\bel{eq:BOint}
\hat{\Bv}_H \approx \;\int_r^{r_J}\,\dint r^\prime \;
\;\frac{\uvr\times\left(\Uv\times\widetilde{\Bv}\right)}{\lambda}
\eep
Since the electrical conductivity profile rules the radial dependence,  
the integral can be approximated by 
\bel{eq:BOintA}
\widehat{\Bv}_H \approx \frac{D_\lambda}{\lambda}
\;\uvr\times\left( \Uv\times\widetilde{\Bv} \right)
\eep
We have assumed here that the LIF vanishes 
at the outer boundary. 
For a dominantly azimuthal flow, the primary LIF component  
is also azimuthal:
\bel{eq:BOintUZ}
\hat{\B}_\phi \approx \frac{D_\lambda}{\lambda}
\;\Uz\tilde{B}_r
\eep
This suggests that the rms value scales with $\RmL$,  
\bel{eq:Best}
\langle\widehat{\Bv}_H\rangle \approx \RmL \langle\widetilde{\Bv}\rangle
\eec
assuming that the correlation between $\Uz$ and $\tilde{B}_r$ is 
of little relevance. 

The radial LIF can be estimated based on the radial
component on the quasi-stationary induction equation \refp{eq:induct}:
\bel{eq:inductr}
\lambda \nabla^2 \hat{B}_r \approx  - \curl\left(\Uv\times\widetilde{\Bv}\right)
\eep
When approximating Ohmic dissipation by 
 $\lambda\hat{B}_r / D_\lambda^2$,  this yields:
\bel{eq:Brest}
\hat{B}_r \approx - \frac{D_\lambda^2}{\lambda}\;\uvr\cdot
 \curl \left( \Uv\times\widetilde{\Bv} \right) 
\eec
which reduces to 
\bel{eq:BrestZ}
\hat{B}_r \approx - \frac{D_\lambda^2}{\lambda}\;
\frac{\Uz}{r\sin{\theta}} \frac{\partial}{\partial\phi} \tilde{B}_r
\ee
for a predominantly  zonal flow. 
This suggest that the rms radial LIF should roughly scale with 
the second modified magnetic Reynolds number 
\bel{eq:Rm2}
 \RmLS = \frac{\langle\Uv\rangle\;D_\lambda^2}{\lambda\;D}
\ee
like 
\bel{eq:BrS}
\langle \hat{B}_r \rangle \approx \RmLS \frac{D}{D_\phi}\;\langle \tilde{B} \rangle
\eep
Here $D_\phi$ is the azimuthal length scale of the background field.
Since $D_\lambda\ll D_\phi$, the radial LIF is 
much smaller than its horizontal 
counterpart \citep{Wicht2019}.

\subsection{Data}

\begin{figure}[h!]
\begin{center}
{\centering
      \includegraphics[width=0.8\columnwidth]{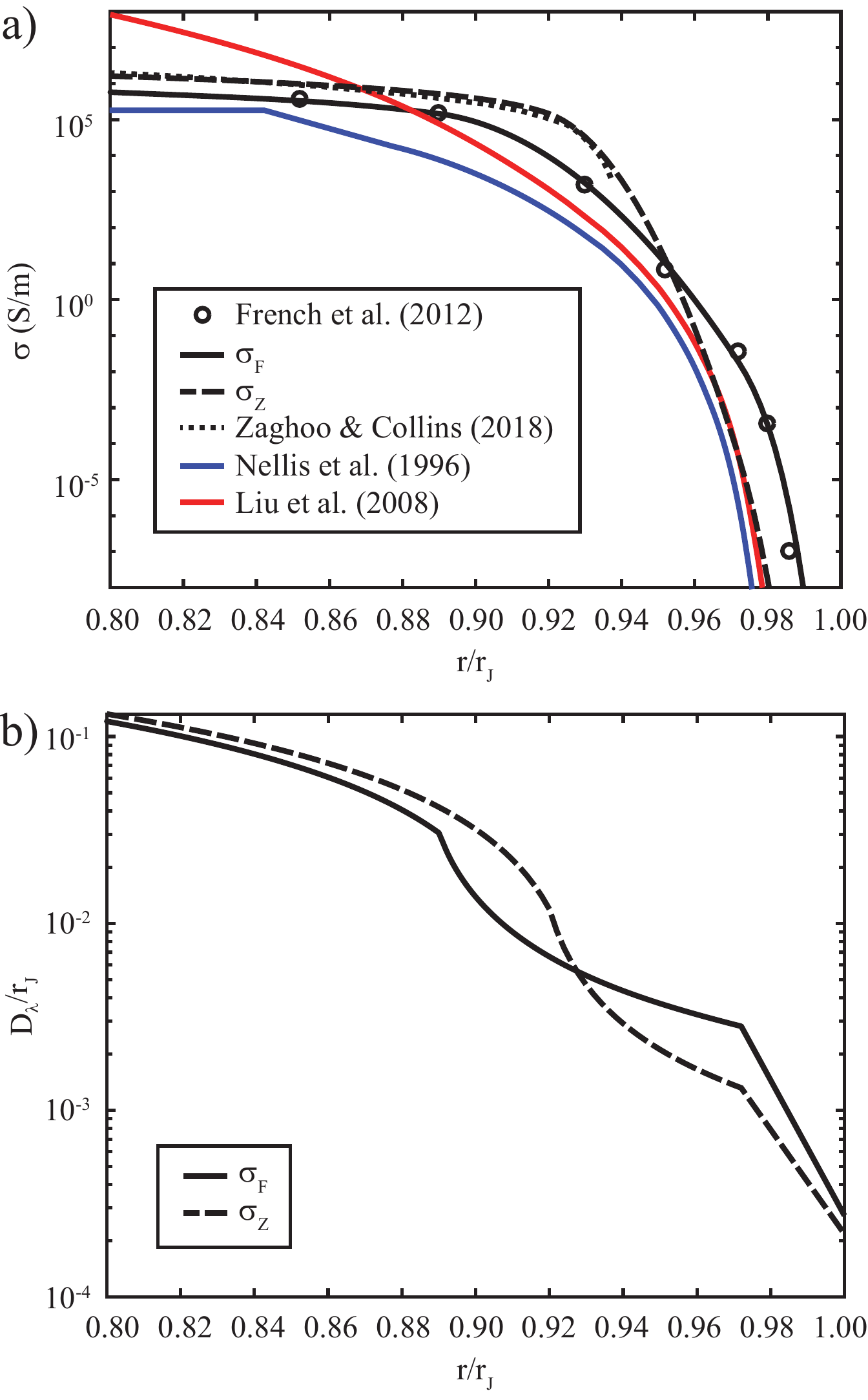}
}
\caption{(a) Electrical conductivity profiles in the outer 20\% of 
Jupiter's radius. The black line shows the parametrization 
$\sigma_F(r)$ of the ab initio simulation data points 
(black circles) by \citet{French2012}. 
The dotted red line shows the profile published 
in \citet{Zaghoo2018}, while the solid red line  
shows the extension $\sigma_Z(r)$ used here. 
The profiles suggested by \citet{Liu2008} (green) 
and \citet{Nellis1999} (blue) are shown for comparison. 
(b) Magnetic diffusivity scale height $D_\lambda/r_J$ 
for $\sigma_F$ and $\sigma_Z$. 
} 
\label{fig:sigma}
\end{center}
\end{figure}

The electric current and LIF estimates discussed above
require a conductivity profile, a zonal flow model, and a surface 
magnetic field model.
For the heating and entropy estimates that we will derived in \secref{sec:Heat}, we 
also need density, temperature, and thermal conductivity 
profiles. We adopt the interior model 
calculated by \citet{Nettelmann2012} and \citet{French2012},
which is the only one proving all the required information. 
Note, however, that recent Juno gravity data suggest 
that Jupiter's interior may be more complex than 
anticipated in this model \citep{Debras2019}.  

Ab initio simulations of the electrical conductivity 
by \citet{French2012} provide 12 data points at 
different depths. \Figref{fig:sigma} shows the values in the 
outer $20$\% of Jupiter's radius and the parametrization $\sigma_F(r)$ 
developed for our analysis. 
A linear branch, 
\bel{eq:cond0}
\sigma_F(r) = \sigma_r + 
(\sigma_m-\sigma_r) \frac{r-r_r}{r_m-r_r}
\eec
covers the smoother inner part $r<r_m$. 
An exponential branch, 
\bel{eq:condG}
\sigma_F(r) =  
\sigma_m \exp\left( 
\left[\frac{r-r_m}{r_m-r_r} + 
      b\,\left(\frac{r-r_m}{r_m-r_r}\right)^2 
\right]\frac{\sigma_m-\sigma_r}{\sigma_m}
\right)
\eec
describes the steeper decay for $r_m<r<r_e$ with
$b=7.2$.  
Matching radius $r_m=0.89\,r_J$ 
and reference radius $r_r=0.77\,r_J$ are chosen where  
ab initio data points have been provided. 

A double-exponential branch,
\bel{eq:condW}
\sigma_F(r) =  
\sigma_e \exp\left( d\,\left[
\exp\left( c \frac{r-r_e}{r_m-r_r} \right) -1 \right] 
 \right)
\eec
is required to capture the super-exponential decrease 
for $r\ge r_e=0.972\,r_J$.
The additional free parameter is $c=10$, 
while $\sigma_e=\sigma(r_e)$ and 
\bel{eq:d}
d = \frac{1}{c}\,\left( 1 + 2\,b\;\frac{r_e-r_m}{r_m-r_i}\right)\;\frac{\sigma_m-\sigma_r}{\sigma_m}
\eep

The dotted red line in \figref{fig:sigma} shows the 
conductivity model used to study dynamo action in Jupiter and 
Jupiter-like exoplanets by \citet{Zaghoo2018}. This is based on 
measurements which suggest a higher electrical conductivity in the 
metallic hydrogen phase than previous data. Unfortunately, \citet{Zaghoo2018}
do not discuss how the results were extrapolated to Jovian 
conditions. The solid red line in \figref{fig:sigma} shows 
the respective parametrization $\sigma_Z(r)$ used for our analysis, which 
retraces the published curve and connects to previously 
published parametrizations (green and blue) 
at lower densities \citep{Nellis1996,Liu2008}. 
Note, however, that these parametrizations are based 
on data \citep{Weir1996} which may have been attributed to 
too low temperatures according to a recent 
analysis by \citet{Knudson2018}. 

Though model $\sigma_Z(r)$ is somewhat arbitrary, it serves to 
illustrate the impact of conductivity uncertainties in our study.
Close to $r_J$ where conductivities remain insignificant, 
$\sigma_F$ is many orders of magnitude larger 
than $\sigma_Z$. The ratio $\sigma_F/\sigma_Z$ 
decreases with depth, reaching
$10^2$ around $0.97\,r_J$ and $10$ around $0.96\,r_J$. 
The two models finally cross at about $0.95\,r_J$.  
At about $0.925\,r_J$, the ratio reaches a minimum of $0.05$ 
and then slowly increases with depth to $0.35$ at $0.8\,r_J$.
\revi{\Tabref{tab:values} list values of both conductivity models
for selected radii.} 

\begin{table*}
{\centering
\begin{tabular}{c|ccccccc}
$r/r_J$ & $0.98$ & $0.97$ & $0.96$ & $0.95$ & $0.94$ & $0.92$ & $0.90$\\
\hline
$\langle U_G\rangle$ [m/s]&$3.0\tp{1}$&$3.2\tp{1}$&$3.5\tp{1}$&
$3.8\tp{1}$&$4.1\tp{1}$&$2.5\tp{1}$&$1.5\tp{1}$\\
$\langle U_K\rangle$ [m/s]&$2.0\tp{1}$&$1.4\tp{1}$&$8.0\tp{0}$&$4.0\tp{0}$&$2.0\tp{0}$&$2.5\tp{-1}$&$4.0\tp{-2}$\\
$\langle U_Z\rangle$ [m/s]&$2.3\tp{1}$&$1.9\tp{1}$&$1.5\tp{1}$&$1.1\tp{1}$&$6.8\tp{0}$&$1.7\tp{0}$&$2.4\tp{-1}$\\
\hline
$\sigma_F\,$[S/m]&$3.2\tp{-4}$&$3.7\tp{-2}$&$9.7\tp{-1}$&$1.7\tp{1}$&
$2.1\tp{2}$&$9.3\tp{3}$&$8.6\tp{4}$\\
$\sigma_Z\,$[S/m]&$1.5\tp{-8}$&$1.9\tp{-4}$&$1.6\tp{-1}$&$3.4\tp{1}$&
$2.0\tp{3}$&$1.5\tp{5}$&$3.9\tp{5}$\\
$\lambda_F\,$[m$^2$/s]&$2.5\tp{9}$&$2.2\tp{7}$&$8.2\tp{5}$&$4.6\tp{4}$&
$3.8\tp{3}$&$8.5\tp{1}$&$9.2\tp{0}$\\
$\lambda_Z\,$[m$^2$/s]&$5.2\tp{13}$&$4.1\tp{9}$&$5.1\tp{6}$&$2.3\tp{4}$&
$3.9\tp{2}$&$5.5\tp{0}$&$2.0\tp{0}$\\
$D_\lambda(\sigma_F)/r_J$&$1.4\tp{-3}$&$2.9\tp{-3}$&$3.2\tp{-3}$&
$3.7\tp{-3}$&$4.4\tp{-3}$&$6.6\tp{-3}$&$1.4\tp{-2}$\\
$D_\lambda(\sigma_Z)/r_J$&$7.9\tp{-4}$&$1.4\tp{-3}$&$1.7\tp{-3}$&
$2.1\tp{-3}$&$2.9\tp{-3}$&$1.2\tp{-2}$&$3.3\tp{-1}$\\
\hline
$\RmL(U_G,\sigma_F)$&$1.2\tp{-3}$&$2.9\tp{-1}$&$9.6\tp{0}$&$2.1\tp{2}$&
$3.2\tp{3}$&$1.4\tp{5}$&$1.5\tp{6}$\\
$\RmL(U_K,\sigma_F)$&$8.2\tp{-4}$&$1.3\tp{-1}$&$2.2\tp{0}$&$2.3\tp{1}$&
$1.6\tp{2}$&$1.4\tp{3}$&$4.2\tp{3}$\\
$\RmL(U_Z,\sigma_F)$&$9.1\tp{-4}$&$1.8\tp{-1}$&$4.2\tp{0}$&$6.2\tp{1}$&
$5.4\tp{2}$&$9.3\tp{3}$&$2.6\tp{4}$\\
$\RmL(U_G,\sigma_F)$&$3.2\tp{-8}$&$7.4\tp{-4}$&$7.9\tp{-1}$&$2.4\tp{2}$&
$2.1\tp{4}$&$3.8\tp{6}$&$1.6\tp{7}$\\
$\RmL(U_K,\sigma_F)$&$2.2\tp{-8}$&$3.2\tp{-4}$&$1.8\tp{-1}$&$2.5\tp{1}$&
$1.0\tp{3}$&$3.9\tp{4}$&$4.3\tp{4}$\\
$\RmL(U_Z,\sigma_F)$&$2.4\tp{-8}$&$4.5\tp{-4}$&$3.4\tp{-1}$&$6.9\tp{1}$&
$3.5\tp{3}$&$2.7\tp{5}$&$2.7\tp{5}$\\
\hline
\end{tabular}
\caption{Rms flow velocities, electrical conductivities $\sigma$, 
magnetic diffusivities $\lambda$, diffusivity scale heights $D_\lambda$, 
and magnetic Reynolds numbers $\RmL$ at selected radii. 
}
\label{tab:values}
}
\end{table*}

Panel b) of \figref{fig:sigma} \revi{and selected 
values listed in \tabref{tab:values}} demonstrate that the magnetic 
diffusivity scale heights $D_\lambda$ differ much less than the conductivities themselves. 
Electric currents, locally induced fields, and Ohmic heating 
depend linearly on $\sigma$ but on different powers 
of $D_\lambda$. The differences between the results for the
two conductivity models is thus predominantly determined by 
$\sigma$ and can easily be scaled from one to the other. 

\begin{figure}[h!]
\begin{center}
{\centering
      \includegraphics[width=0.8\columnwidth]{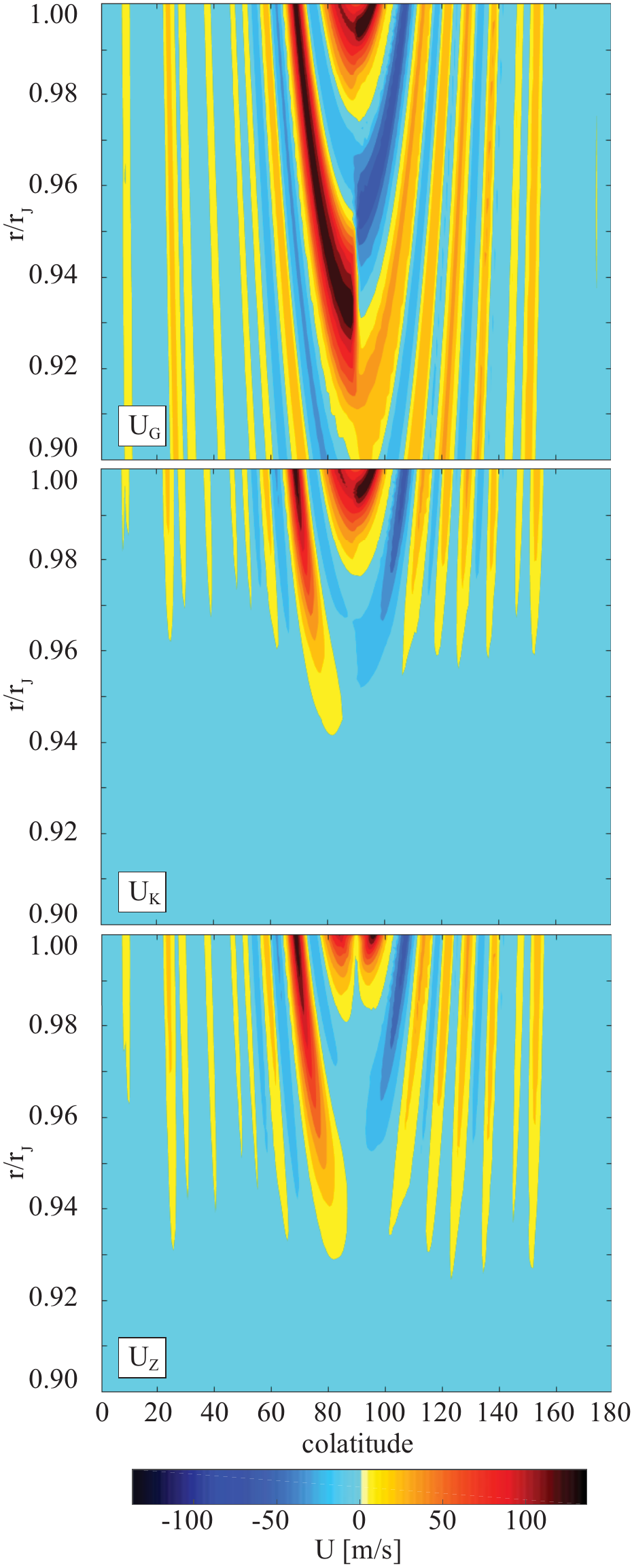}
}
\caption{Zonal flow models used in this study. Prograde flows are shown in 
red and yellow, while blue indicates retrograde directions.}
\label{fig:UZ}
\end{center}
\end{figure}

The three different zonal flow models explored here are  
illustrated in \figref{fig:UZ}. 
\revi{\Tabref{tab:values} lists rms values $\langle \Uz\rangle$ at 
selected radii.} 
All reproduce the observed zonal winds at $r=r_J$  
\citep{Porco2003,Vasavada2005}. 
\revi{We use running averages of the surface profiles
with a window width of one degree and represent the result  
with 256 (nearly) evenly spaced latitudinal grid points
for our calculations.} 

The three flow models differ at depth. 
The most simple one, $U_G$,  
assumes geostrophy in each hemisphere, 
i.e.~the flow depends only on the distance $s=r \sin\theta$ 
to the rotation axis. 
\citet{Kaspi2018} describe 
the depth decay of the equatorially antisymmetric zonal flow
with profiles constrained by the Juno gravity measurements. 
We apply their 'latitude independent' model version 
to the total zonal flow and refer to this model as $U_K$. 
The rms amplitude of $U_K$   
has decreased by one order of magnitude at about $0.95\,r_J$
and by two orders of magnitude around $0.925\,r_J$.  

We also consider the 'deep' model suggested 
by \citet{Kong2018}, who assume an exponential depth decay 
and an additional linear dependence on the distance $z=\cos\theta$ 
to the equatorial plane. Like for $U_K$, our respective model $U_Z$ 
assumes that the depth and $z$ dependencies, which were originally derived 
for the equatorially antisymmetric contributions, apply to 
the whole flow. The rms velocity in $U_Z$ decays smoother 
with depth than in $U_K$, having decreased by one 
order of magnitude at about $0.935\,r_J$ and by two orders 
of magnitude at about $0.905\,r_J$. 

\Figref{fig:UZ} shows that $U_G$ and $U_K$ 
have discontinuities at the equatorial plane. 
\revi{These pose a problem when calculating 
the latitudinal zonal flow derivatives 
required for the integral estimate 
$j^{(I)}_\theta$ (see \eqnref{eq:JItO}). 
Formally, the derivative becomes infinite at the equator. 
Practically, however, the impact of the discontinuity 
depends on the model setup and on the methods used 
for calculating the derivatives. 
We tested the impact on rms current density estimates 
by comparing calculations covering all latitudes 
with counterparts where the derivatives were explicitly 
set to zero in a six-degree belt around the equator. 
Simple first order finite differences with 256 grid 
points at each radial level are generally used for calculating
the derivative. 
For flow $U_Z$, which has been constructed to avoid the 
discontinuity \citep{Kong2018}, the belt contributes not more than 
one percent to $\langle j^{(I)}_\theta\rangle$ at any radius,
which is less than the surface fraction it represents. 
For flow $U_K$, the contribution is even smaller 
due to the faster decay of the flow amplitude. 
However, for $U_G$ the belt contributes $20$\% to
the rms current for radii below $0.94\,r_J$, 
which is a clear sign that the unphysical 
discontinuity causes problems. 
In order to be on the safe side, 
we will only consider  flow model $U_Z$ 
in connection with estimate $j^{(I)}_\theta$ below.}

\begin{figure}[h!]
\begin{center}
{\centering
      \includegraphics[width=0.9\columnwidth]{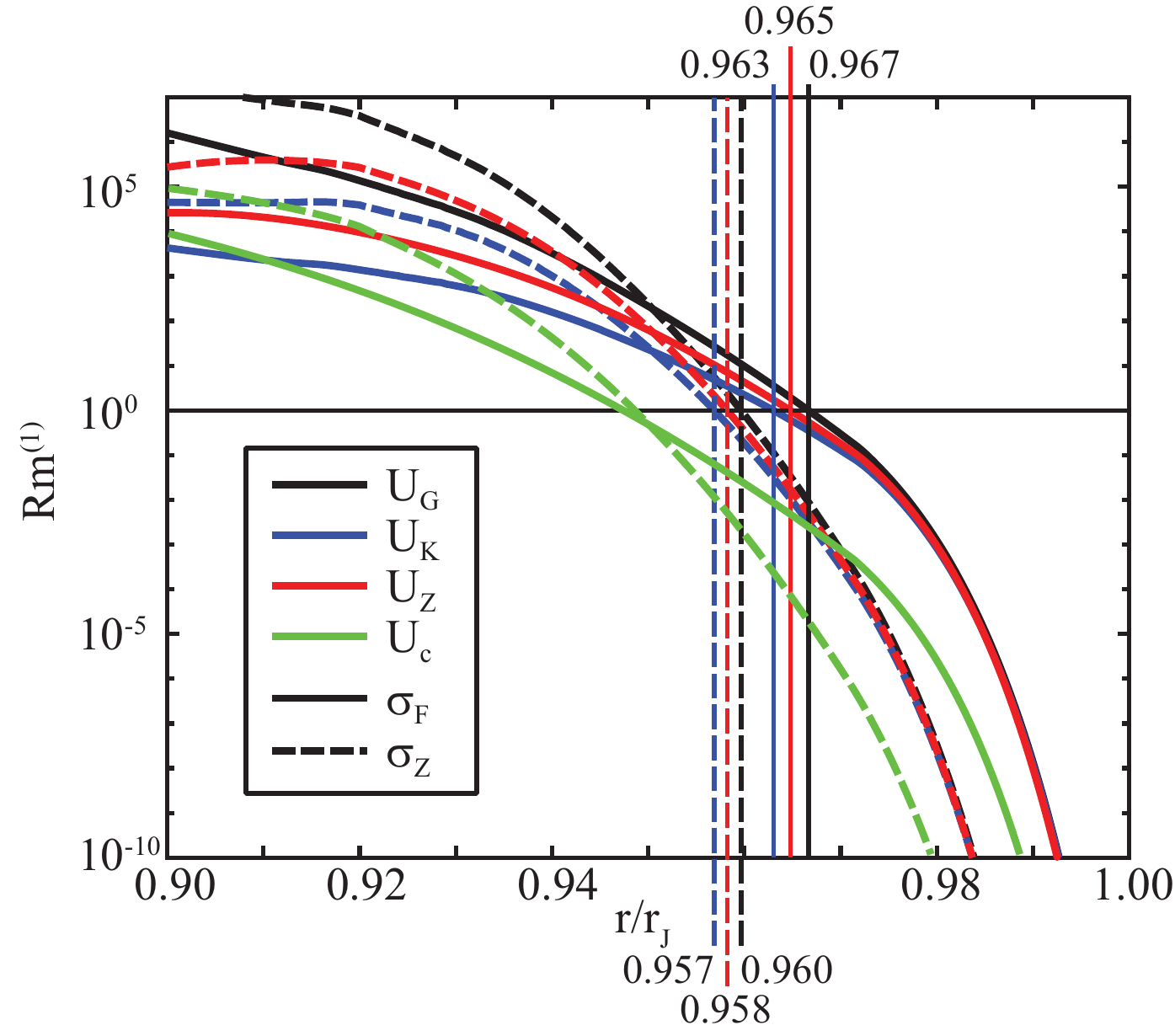}
}
\caption{Modified magnetic Reynolds numbers for the 
three zonal flow and the two conductivity models explored here. 
Vertical lines mark where the radii $r_1$ where $\RmL=1$.
Green lines show profiles for a typical convective velocity 
of $10\,$cm/s suggested by scaling laws. 
} 
\label{fig:RmL}
\end{center}
\end{figure}

The radius where $\RmL=1$, which we will refer to as 
$r_1$ in the following, roughly marks the point 
where the approximations discussed above break down \citep{Wicht2019}. 
\Figref{fig:RmL}a)  
illustrates the $\RmL$ profiles 
that result from combining $\sigma_F$ and $\sigma_Z$ 
with rms values for the three zonal flow models.
\revi{\Tabref{tab:values} compares values at 
some selected radii.}
These modified magnetic Reynolds numbers 
exceed unity between $r_1=0.957\,r_J$ 
for the combination $\sigma_Z$ and $U_K$ and  
$r_1=0.967\,r_J$ for $\sigma_F$ and $U_G$. 
All $r_1$ values are listed in \tabref{tab:crossing}.

Green lines in \figref{fig:RmL} show $\RmL$ profiles 
for a typical convective velocity of $10\,$cm/s 
suggested by scaling laws \citep[e.g.~see ][]{Duarte2018}. 
Numerical simulations show that the velocity increases with 
radius, an effect not taken into account here. The comparison of the 
different $\RmL$ profiles suggests that zonal flow related 
dynamo action should dominate at least in the outer 
$9$\% in radius. 

For Jupiter's surface magnetic field we use the JRM09 model by 
\citet{Connerney2018}, which provides information 
up to spherical harmonic degree $\ell=10$. 
The more recent model by \citet{Moore2018} 
is only slightly different. 
In order to check the impact of smaller scale  
contributions, we also tested the numerical model G14 
by \citet{Gastine2014}, 
which reproduces Jupiter's large scale field and provides 
harmonics up $\ell=426$. 
Since it turned out that the impact of the smaller scales is
very marginal, the results are not shown here. 

\section{Dissipative Heating and Entropy Production in Jupiter}
\label{sec:Heat}

\citet{Liu2008} constrain the depth of the 
zonal winds by assuming that the related total Ohmic 
heating should not exceed the heat flux 
out of the planet. Unfortunately, 
this assumption is not correct, as we will show in the 
following. In order to arrive 
at more meaningful constraints, we start with reviewing 
some fundamental considerations. 

In a quasi-stationary state, where flow and magnetic field 
are maintained by buoyancy and induction against 
dissipative losses, the conservation of energy simply 
states that the heat flux $Q_o=Q(r_o)$ through the outer 
boundary is the sum of the flux $Q_i=Q(r_i)$ through the 
inner boundary and the total internal heating $H$: 
\bel{eq:EB}
Q_o = Q_i + H
\eep
Note that neither viscous nor Ohmic heating contribute to $H$. 
Since flow and magnetic field are maintained by the heat flux through 
the system, they cannot be counted as net heat  
sources \citep{Hewitt1975,Braginsky1995}. 

When furthermore also neglecting the effects of helium segregation, 
core erosion, or planetary shrinking as potential 
energy sources, the only remaining contribution 
is the slow secular cooling of the planet. 
The volumetric heat source is then given by 
\bel{eq:h}
 h = \rhob\,\Tb\, \frac{\partial \tilde{S}}{\partial t}
\eec
where the tilde indicates the hydrostatic, adiabatic, background state 
\citep{Braginsky1995}.
Assuming that convection maintains an adiabat at all times, 
$\partial\tilde{S} / \partial t$ remains homogeneous throughout 
the convective region and obeys \citep{Jones2014}: 
\bel{eq:C}
\frac{\partial \tilde{S}}{\partial t} = 
\left(Q_o-Q_i\right)\big/\int_V \dint V\; \rhob\,\Tb
\eep 
Here, $\int_V \dint V$ denotes an integration over
the whole convective volume.  
Note, however, that the thermal evolution could be 
more complex, should Jupiter indeed harbor stably stratified regions. 

In order to get a handle on dissipative heating, one has to 
consider the local heat equation 
\bel{eq:FEH}
\rhob \Tb \left(\frac{\partial s}{\partial t} + \Uv\cdot\nabla s \right) 
= \nabla\cdot\left(k \nabla T\right) + h + \dissi
\eec
where $\dissi$ denotes the volumetric dissipative heat source, 
and $k$ is the thermal conductivity. 
When assuming a steady state and adopting the anelastic 
approximation $\nabla\cdot(\rhob\Uv)=0$, the 
integration over the shell between the inner boundary $r_i$ and 
radius $r$ yields 
\bel{eq:FEHR}
Q_D(r) + Q_A(r)  =  
Q_i  + \int_{r_i}^r \dint r^\prime \int_F \dint F\;\left( h + \dissi + 
\rhob\;s\;U_r \frac{\partial \Tb}{\partial r} \right)
\eep
The left hand side is the 
total flux through the boundary at $r$, i.e.~the sum 
of the diffusive contribution 
\bel{eq:Qr}
 Q_D(r) = - \int_F \dint F\;k \frac{\partial T}{\partial r}
\ee
and the advective contribution 
\bel{eq:QA}
Q_A(r) = \int_{F(r)} \dint F\;\rhob\, \Tb\,s\,U_r
\eep
The right hand side of \eqnref{eq:FEHR}  
reflects the influx through the lower boundary $Q_i$ plus 
three volumetric contributions: the slow secular cooling, 
the dissipative heating, and the adiabatic cooling. 
Writing the adiabatic cooling in terms of $Q_A$ yields the relation
\beal{eq:FEHR2}
Q_D(r) + Q_A(r)  = 
Q_i  & + & \int_{r_i}^r \dint r^\prime \int_F \dint F\;\left( h + \dissi \right) \\ &-&\nonumber  
\int_{r_i}^r \dint r^\prime Q_A(r^\prime)\,\big/\, D_T(r^\prime)\;\;,
\eea
where $D_T=-\tilde{T}/(\partial \tilde{T}/\partial r)$ is the thermal scale height.

Integrating \eqnref{eq:FEHR2} over the whole convective volume 
and using \eqnref{eq:EB} reveals that the total dissipative 
heating $\Dissi_T$ is balanced by the total adiabatic cooling:
\bel{eq:AC}
\Dissi_T = \int_V \dint V \;\dissi =  \int_{r_i}^{r_o} \dint r^\prime\; Q_A(r^\prime)
\,\big/\, D_T(r^\prime)
\eep
The total adiabatic cooling is actually identical to the buoyancy power $P$ 
that drives convection and thus the dynamo mechanism.
Multiplying the buoyancy term in the Navier-Stokes equation 
with velocity and integrating over the convective volume 
to yield the total convective power input indeed 
gives the same expression \citep{Braginsky1995}.
\Eqnref{eq:AC} thus simply states that dissipation is balanced by 
the power input $P$ to the system, a fact used in many 
scaling laws to establish how the rms magnetic field strength 
or rms velocity scale with $P$ \citep{Christensen2006,Christensen2009,Davidson2013,Yadav2013}. 

\Eqnref{eq:AC} requires to know $Q_A$ at 
each radius. Since $Q_A$ itself depends on the 
distribution of dissipative heat sources, however, 
an additional condition is required. Assuming that Ohmic 
heating and adiabatic cooling not only cancel globally but,
at least roughly, also at each radius offers a simple solution
used in most scaling laws (though never stated explicitely). 
With the exception of thin thermal boundary layers, the 
heat flux is then dominated by the advective contribution, 
so that 
\bel{eq:QAA}
Q_A(r) \approx   Q_i +   \int_{r_i}^r \dint r \int_F \dint F\; h
\eep

Adopting the interior model by \citet{Nettelmann2012} and 
\citet{French2012} and 
the observed flux $Q_o=3.35\tp{17}\,$W  
from the planet's interior \citep{Guillot2015} 
allows calculating $h$ via
\eqnref{eq:h}. Because the inner core occupies 
only $10$\% in radius, $Q_i$ can be neglected. 
When, for example, assuming that $h$ also 
describes the cooling of the rocky core, $Q_i$ is two orders of
magnitude smaller than $Q_o$. 

Plugging \eqnref{eq:QAA} into \eqnref{eq:AC} finally 
allows calculating \revi{the total dissipative heating}:
\bel{eq:PV}
\Dissi_T = 1.20\tp{18}\,\mbox{W}
\eep
The result reveals that dissipative heating can in fact exceed the 
heat flux out of Jupiter's interior by a factor of $3.6$.
\citet{Gastine2014} came up with a power estimate that is 
about $50$\% smaller because they used 
a simplified formula provided by \citet{Christensen2009}. 

Considering the entropy rather than the heat balance avoids  
the need to come up with an additional condition 
\citep{Hewitt1975,Gubbins1979,Braginsky1995,Gubbins2003}. 
Dividing the heat equation \eqnref{eq:FEH} by temperature and 
integrating over the convective volume yields the entropy budget 
\bel{eq:ENT}
\frac{Q_o}{\Tb_o} = \frac{Q_i}{\Tb_i} \;+\;
\int_V \dint V \left( \frac{h}{\Tb} \;+\; k\;\left|\frac{\partial T / \partial r}{\Tb}\right|^2 
\;+\; \frac{\dissi}{\Tb} \right)
\eec
where we have once more 
used the anelastic approximation $\nabla\cdot(\rhob\Uv)=0$.
When assuming that the temperature profile stays close to the adiabat, 
the total dissipative entropy production $\varTheta$ can thus 
be approximated by: 
\bel{eq:ent}
\Ent_T = \int_V \dint V \frac{\dissi}{\Tb} \approx \frac{Q_o}{\Tb_o} - \frac{Q_i}{\Tb_i}
 -  \int_V \dint V \frac{h}{\Tb} - 
\int_V \dint V k\;\left|\frac{\partial \Tb / \partial r}{\Tb}\right|^2 
\eep

An upper bound for the total dissipative heating can be derived 
when assuming that $\Tb_i$ is the highest temperature in the system 
\citep{Hewitt1975,Currie2017}:
\bel{eq:best}
\Dissi\;<\; \Tb_i \int_V \dint V \frac{\dissi}{\Tb}\;<\; 
\frac{\Tb_i}{\Tb_o}\; Q_o
\eep
Using once more the internal model by \citet{Nettelmann2012} puts the 
upper bound at $10^2\,Q_o$ for Jupiter, which is at least consistent 
with estimate \refp{eq:PV}.   

When complementing the internal model with    
the thermal conductivity profile by \citet{French2012}, 
we can quantify the different terms in Jupiter's  
entropy budget \refp{eq:ent}. 
Because of the strong temperature contrast between the outer 
boundary and the deeper convective region, 
the entropy flux through the outer boundary clearly dominates.
\revi{The total dissipative entropy production is thus given by:}
\bel{eq:EPM}
\Ent_T \approx Q_o/\tilde{T}_o = 2.0\tp{15}\,\mbox{W/K}
\eep 
The second largest term in \eqnref{eq:ent}, 
the entropy due to the secular cooling, 
is already two orders of magnitude smaller at $3.0\tp{13}\,$W/K.  
The two remaining terms, entropy flux through the inner 
boundary and the diffusive entropy flux down the adiabat, 
are only of order $10^{11}\,$W/K. 

Since the magnetic diffusivity is about $10^6$ times larger than 
its viscous counterpart in planetary dynamo regions, Ohmic heating  
by far dominates. 
We can use the current density estimates to predict the Ohmic 
heating due to the zonal flows above radius $r$: 
\bel{eq:DISR}
\Dissi_O(r) = \int_r^{r_J}\;\dint r^\prime\;\int_F \dint F\;
\frac{\jv^2}{\sigma}
\eep
The conditions
\bel{eq:DissiTotC}
\Dissi_O(r) \le \Dissi_T =  1.20\tp{18}\,\mbox{W}
\ee
provides a possible constraint for the depth of the zonal winds 
in Jupiter.

The dissipative entropy production related to the Ohmic heating 
is given by 
\bel{eq:ENTR}
\Ent_O(r) \approx 
\int_r^{r_J}\;\dint r^\prime\;\int_F \dint F\;
\frac{\jv^2}{\sigma\,\Tb}
\eep  
This can be used for the alternative depth constraint  
\bel{eq:EntC}
\Ent_O(r) \le \Ent_T = 2.0\tp{15}\,\mbox{W/K}
\eep
\section{Dynamo Action in Jupiter's SDCR}
\label{sec:Heating}

\subsection{Electric Currents and Locally Induced Field}

\begin{figure}
\centering
\includegraphics[draft=false,width=0.9\columnwidth]{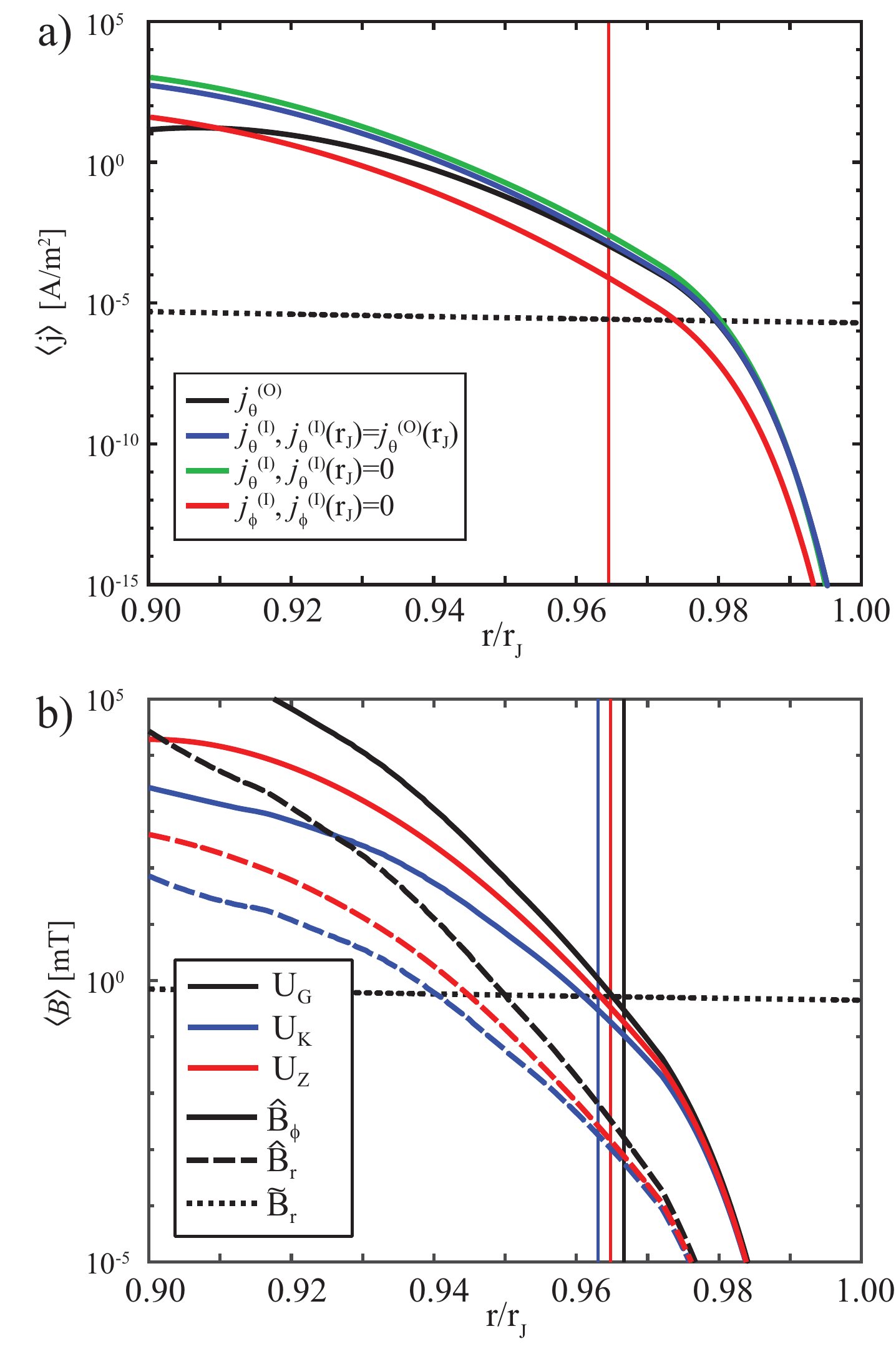}
\caption{(a) Rms current density estimates 
\revi{for flow model $U_Z$ and conductivity model $\sigma_F$}.  
(b) Estimates of rms radial 
and horizontal LIF. The profile of $\langle\tilde{B}_r\rangle$ 
has been included for comparison. Vertical lines 
mark the radii where $\RmL=1$ for the different flow models.
Current density estimates $\jv^{(O)}$ \revi{and 
conductivity model $\sigma_F$} have been used.  
}
\label{fig:JB}
\end{figure} 

We start with discussing the current estimates for the different zonal flow 
and conductivity model combinations. 
\Figref{fig:JB}a) compares rms values of 
integral estimates $\jv^{(I)}$ 
and Ohm's law estimates $\jv^{(O)}$ for 
conductivity model $\sigma_F$ and flow $U_Z$. 
The integral estimates of the latitudinal currents are at least 40 
times larger than their azimuthal counterparts 
for all conductivity and flow model combinations.
When using $\jv^{(O)}$ as outer boundary 
condition for $\jv^{(I)}$  
(blue line in \figref{fig:JB}), 
both estimates remain very similar down to about $0.98\,r_J$. 
At $0.97\,r_J$, however, $\jv^{(I)}$ is already about $50$\% larger 
than $\jv^{(O)}$, and at $0.96\,r_J$ the 
difference has increased to about $250$\%. 
When assuming a vanishing outer boundary current for $\jv^{(I)}$, 
the differences are even larger: 
$\jv^{(I)}$ (green line) is $3.5$ 
times larger than $\jv^{(O)}$ at $0.97\,r_J$ 
and about $6$ times larger at $0.96\,r_J$. 

Estimates of the rms horizontal and radial LIF 
are shown in \figref{fig:JB}b),   
based on $\jv^{(O)}$ and  
\eqnref{eq:BOint} for the horizontal and on \eqnref{eq:BrestZ} 
for the radial components. 
The radial LIF is between two and three orders of magnitude 
smaller than its horizontal counterpart. 
The rougher estimates \refp{eq:Best} and 
\refp{eq:BrS}, based on $\RmL$ and $\RmLS$ respectively, 
provide values that are less than a factor two smaller 
and can thus safely be used for order of magnitude assessments. 
They correctly predict that the rms azimuthal LIF reaches 
the level of the background field at $r_1$  
and also that the ratio of radial to azimuthal LIF 
is about $\RmLS / \RmL=D_\lambda/r_J$. 
At $r_1$, the rms radial LIF is thus roughly  
three orders of magnitude smaller than the background field 
or the horizontal LIF. 
\Tabref{tab:crossing} lists the relative rms radial LIF (column 7) 
at $r_1$ (column 3) 
for all $\sigma$ and flow combinations when using $\jv^{(O)}$.  

\begin{table*}
{\centering
\begin{tabular}{ccccccccc}
conduct. & flow & $r_1\big/r_J$ &  $r_{10}\big/r_J$ & $r_\Dissi\big/r_J$ & $r_\Ent\big/r_J$  & 
$\langle\hat{B}_r\rangle\big/ \langle\tilde{B}_r\rangle$ & $\Dissi_O\big/\Dissi_T$ &  $\Ent_O\big/\Ent_T$\\
\hline
$\sigma_F$ & $U_G$ & $0.967$  & $0.960$ & $0.961$ & $0.955$ & $2.8\tp{-3}$  & $1.2\tp{-1}$ & $2.3\tp{-2}$ \\
                  & $U_Z$ & $0.965$  & $0.957$ & $0.956$ & $0.948$ & $2.2\tp{-3}$  & $8.0\tp{-2}$ & $1.5\tp{-2}$ \\
                  & $U_K$ & $0.963$  & $0.954$ & $0.946$ & $0.929$ & $3.0\tp{-3}$  & $4.8\tp{-2}$ & $9.0\tp{-3}$ \\
\hline
$\sigma_Z$ & $U_G$ & $0.960$  & $0.956$ & $0.956$ & $0.953$ & $1.8\tp{-3}$ & $1.5\tp{-1}$ & $2.5\tp{-2}$ \\
                  & $U_Z$ & $0.958$  & $0.954$ & $0.953$ & $0.949$ & $1.6\tp{-3}$ & $8.9\tp{-2}$ & $1.5\tp{-2}$ \\
                  & $U_K$ & $0.957$  & $0.952$ & $0.948$ & $0.942$ & $1.7\tp{-3}$ & $2.9\tp{-2}$ & $4.9\tp{-3}$ \\
\hline
\end{tabular}
\caption{While the first two columns define the model, columns 3 to 6 list different radii: radius 
$r_1$ where $\RmL=1$, $r_{10}$ where $\RmL=10$, $r_\Dissi$ where $\Dissi_O=\Dissi_T$, and $r_\Ent$ where $\Ent_O=\Ent_T$. 
Column 7 provides the ratio of rms radial LIF to rms radial background field at $r_1$. 
Columns 8 and 9 give ratios $\Dissi_O/\Dissi_T$ and $\Ent_O/\Ent_T$ at $r_1$,  respectively.  
Current density estimates $\jv^{(O)}$ have been used. 
}
\label{tab:crossing}
}
\end{table*}

\begin{figure}
\centering
\includegraphics[draft=false,width=0.95\columnwidth]{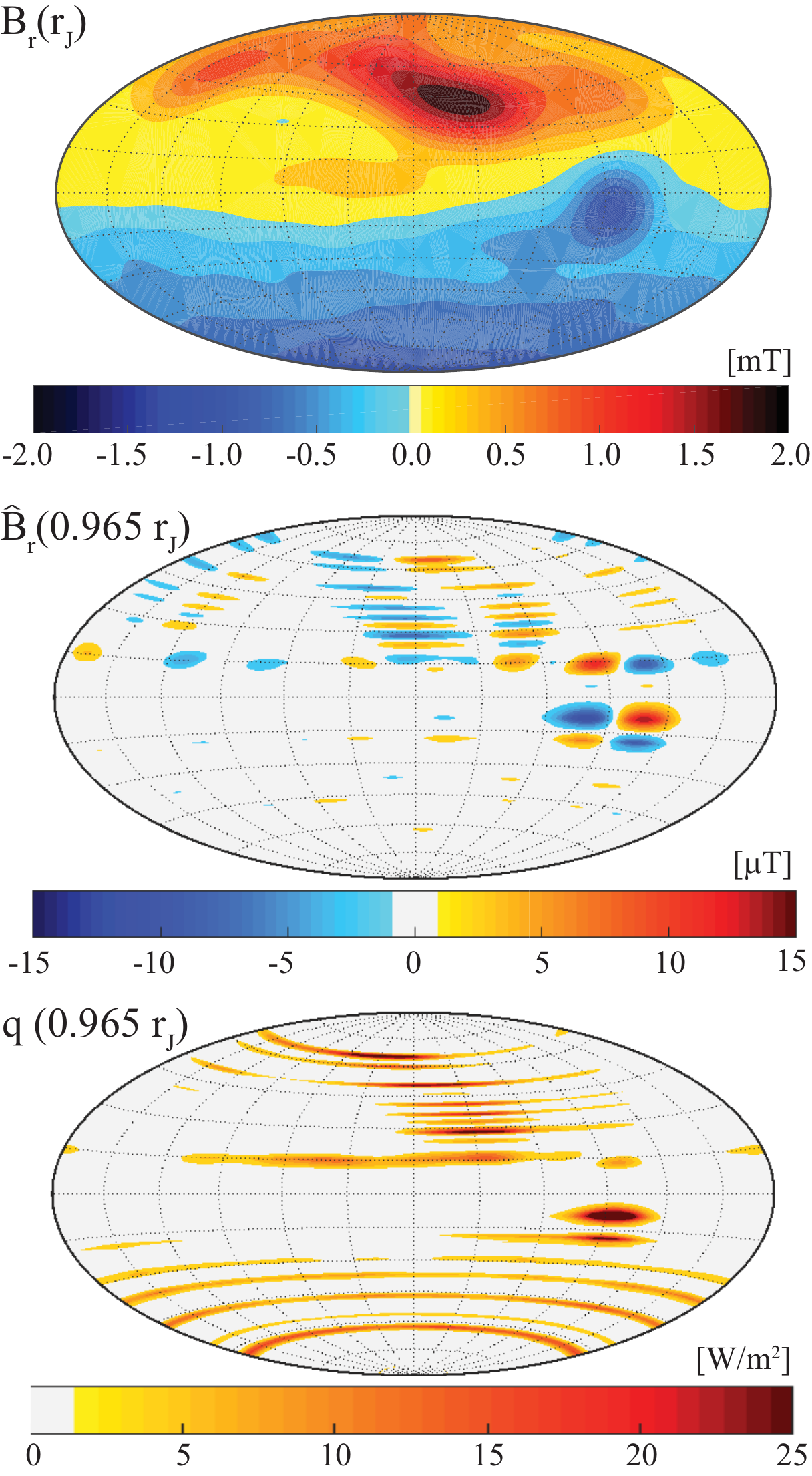}
\caption{Maps of (a) the radial surface field in the 
Jupiter field model JRM09, 
(b) the radial LIF at $r_1=0.965\,r_J$, 
and (c) the local Ohmic heating 
$q=\int_r^{r_J} d r\,\jv^2/\sigma$ at $r=r_1=0.965\,r_J$. 
Flow model $U_Z$, conductivity $\sigma_F$, and 
current density estimates $\jv^{(O)}$ (for panel c) 
have been used.  
Outward (inward) directed field is shown in red (blue).
} 
\label{fig:Maps}
\end{figure} 

\citet{Wicht2019} demonstrate that the Ohm's-law based 
estimates not only provide good rms but also decent local values
for their Jupiter-like dynamo simulations.
\Figref{fig:Maps} shows the radial surface field of 
JRM09 in panel a) and  the radial LIF 
for $\sigma_F$ and $U_Z$ at $r_1$ in
panel b). 
A very distinct pattern of localized field patches can 
be found where the fast zonal jets around the equator 
interact with the strong blue patch in the JRM09 model. 

\begin{figure}
\centering
\includegraphics[draft=false,width=0.9\columnwidth]{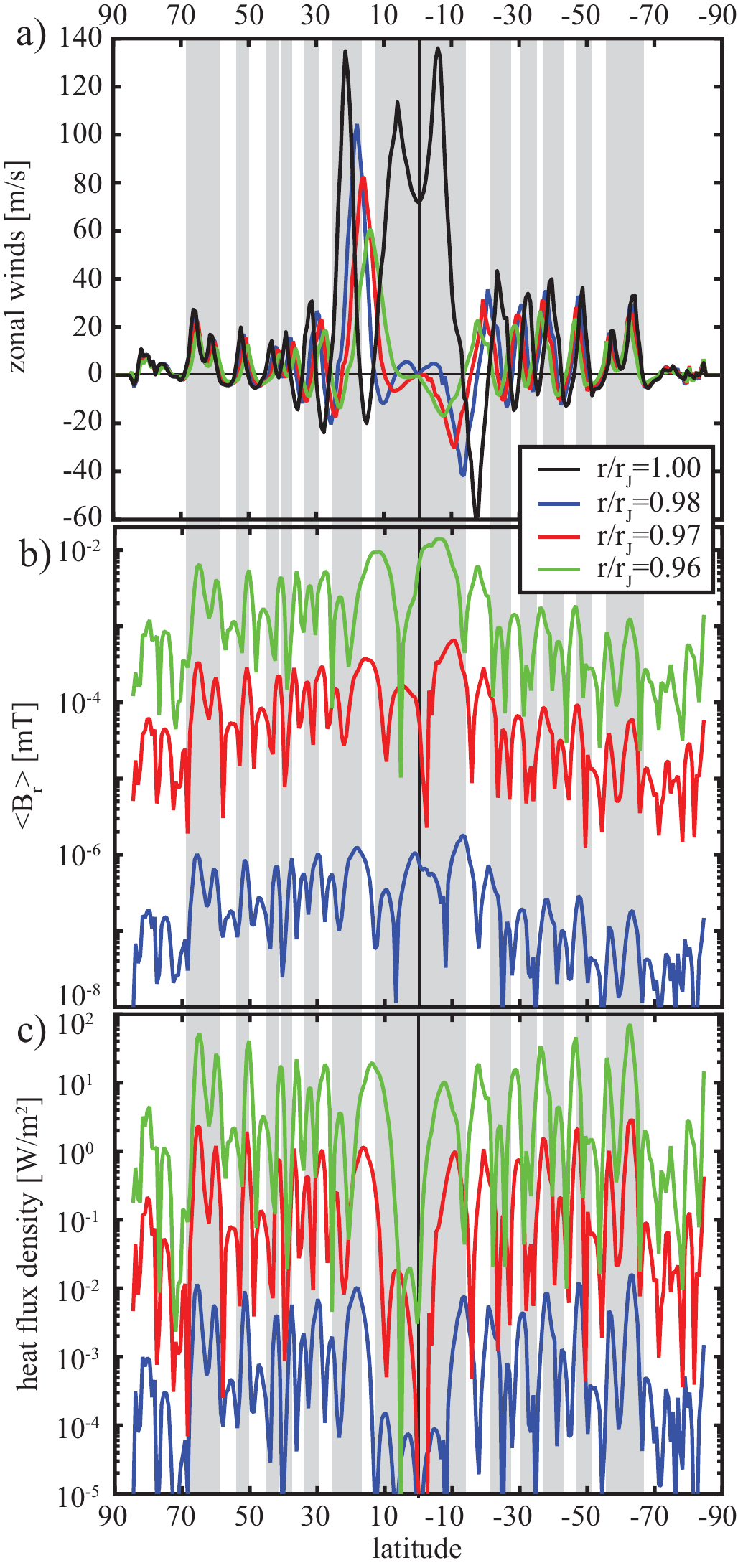}
\caption{(a) $U_Z$ profiles at different depths; 
the grey background color highlights 
where surface jets are prograde. 
(b) azimuthal rms radial LIF and (c) 
heat flux $q=\int_r^{r_J} d r\,\jv^2/\sigma$ at three different depths.  
Conductivity model $\sigma_F$, flow $U_Z$, and estimate $\jv^{(O)}$ have been used. 
} 
\label{fig:AS}
\end{figure} 

The zonal flow pattern remains recognizable in the LIF, 
as is clearly demonstrated in \figref{fig:AS}, which compares 
zonal flow profiles in panel a) with the 
azimuthal rms of the radial LIF in panel b). 
Due to the flow geometry, the 
currents and LIF show a depth-dependent phase shift relative to the surface jets.  
The equatorial jet, which is so prominent at the surface, contributes very little to dynamo action, 
since it does not reach down to depths where the electrical
conductivity is more significant. 

\begin{figure}
\centering
\includegraphics[draft=false,width=0.9\columnwidth]{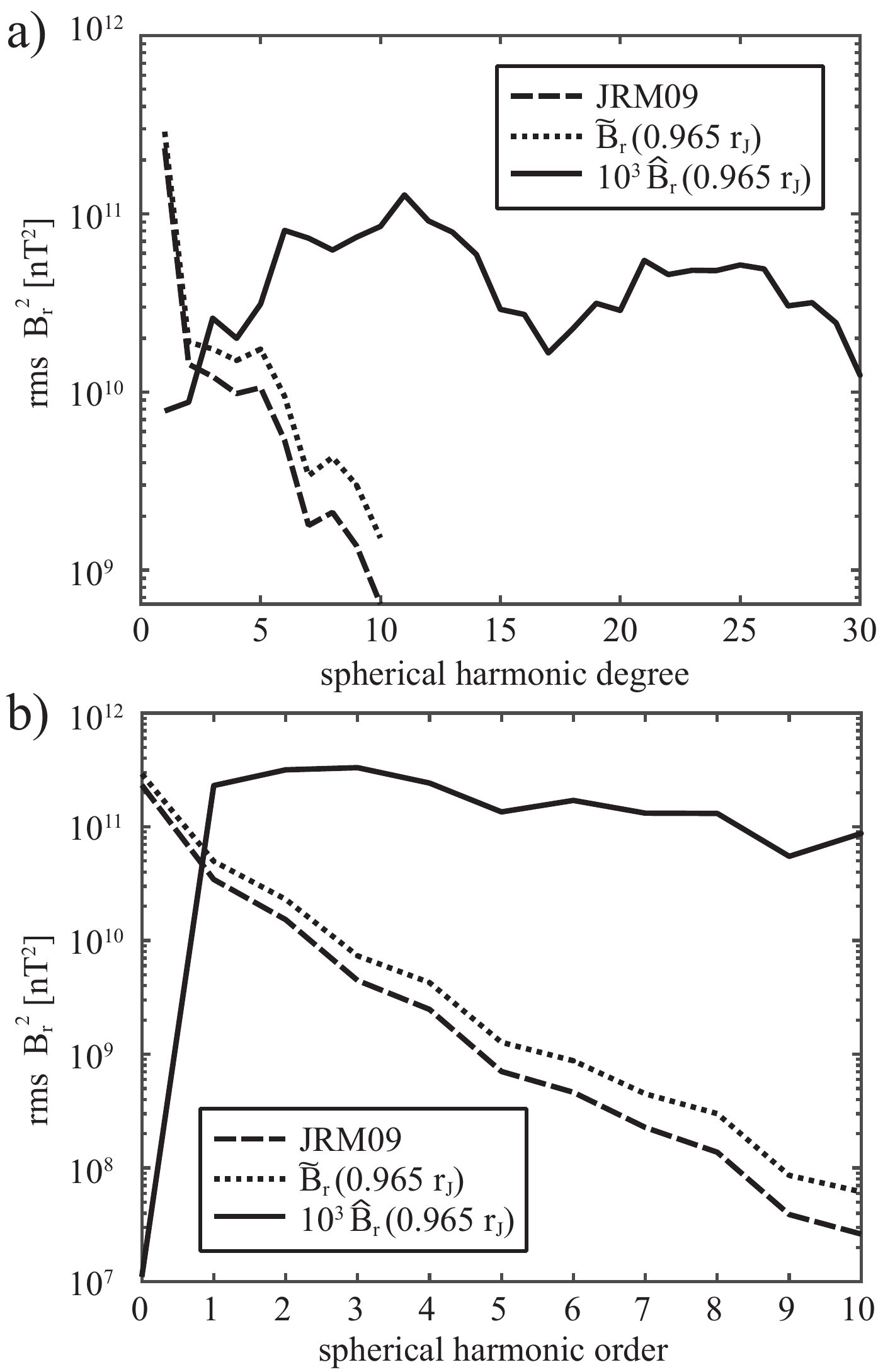}
\caption{Power spectra of rms radial field contributions
 (Mauersberger-Lowes) for JRM09, the downward continued 
$\tilde{B}_r$  and the radial LIF $\hat{B}_r$ 
at $r_1=0.965\,r_J$. \revi{(a) shows the spherical harmonic 
degree spectrum, while (b) shows the harmonic order spectrum.} 
The LIF has been amplified with $10^3$. 
Flow $U_Z$ and conductivity $\sigma_F$ have been used.  
} 
\label{fig:Spectra}
\end{figure} 

\Figref{fig:Spectra} compares spherical harmonic power spectra
of the background radial field and the radial LIF. As already 
apparent from the map shown in \figref{fig:Maps}, the LIF 
is dominated by smaller scale contributions. The spherical
harmonic degree spectrum results from the convolution of 
the complex latitudinal zonal flow structure with the background field. 
At $r_1$, 
the dipole contribution in the LIF is about $10^{-4}$ times smaller
than the respective background field contribution. 
For degree $\ell=10$, the ratio has increased to $10^{-2}$.   
The spectrum peaks at $\ell=12$ but has also 
significant contributions from even higher degrees. 

The spherical harmonic order spectrum, shown in 
panel b) of \figref{fig:Spectra}, is very different. 
The action of the axisymmetric zonal flow on $\tilde{B}_r$ 
excites no additional harmonic orders so that 
the spectrum remains confined to $m\le10$.
The LIF spectrum is rather flat but has no 
axisymmetric contribution. At $m=10$, 
the rms LIF amplitude reaches roughly $25\,$\% of the 
background field.

The results for the conductivity model $\sigma_F$ presented so far
can roughly be scaled to model $\sigma_Z$ by 
multiplying with the conductivity ratio $\sigma_Z/\sigma_F$. 
Around $0.97\,r_J$, the LIF is two orders of magnitude 
weaker, and the difference decreases with depth, reaching about one order 
of magnitude around $0.96\,r_J$. 
Where $\RmL=1$, on the other hand, the LIF reaches comparable 
values for both conductivity models.
The different flow models yield very similar LIF 
pattern, albeit with the different amplitudes 
indicated in \figref{fig:JB}.

\subsection{Ohmic Heating and Entropy Constraint}

We now use the electric current estimates to calculate 
Ohmic heating and entropy production.
Panel c) of \figref{fig:Maps} shows the map 
of Ohmic heat flux density $q=\int_r^{r_J} d r\,\jv^2/\sigma$ 
at radius  $r_1$ when using $\jv^{(O)}$, $\sigma_F$, and $U_Z$.
The currents induced by interaction between the fierce zonal 
jets close to the equator and the strong blue patch 
in JRM09 not only yield a highly localized LIF but also intense 
local heating. While the action of various other zonal jets reaches 
a lower level, the related pattern remains roughly recognizable in the form 
of thin heating bands. The azimuthal mean of $q$, shown in 
Panel c) of \figref{fig:AS}, clearly illustrates 
the correlation between heating and the zonal jets. 
Like for the LIF, there is a depth-dependent phase 
shift between the observed surface zonal wind profile 
and the Ohmic heating pattern. 

\begin{figure}
\centering
\includegraphics[draft=false,width=0.9\columnwidth]{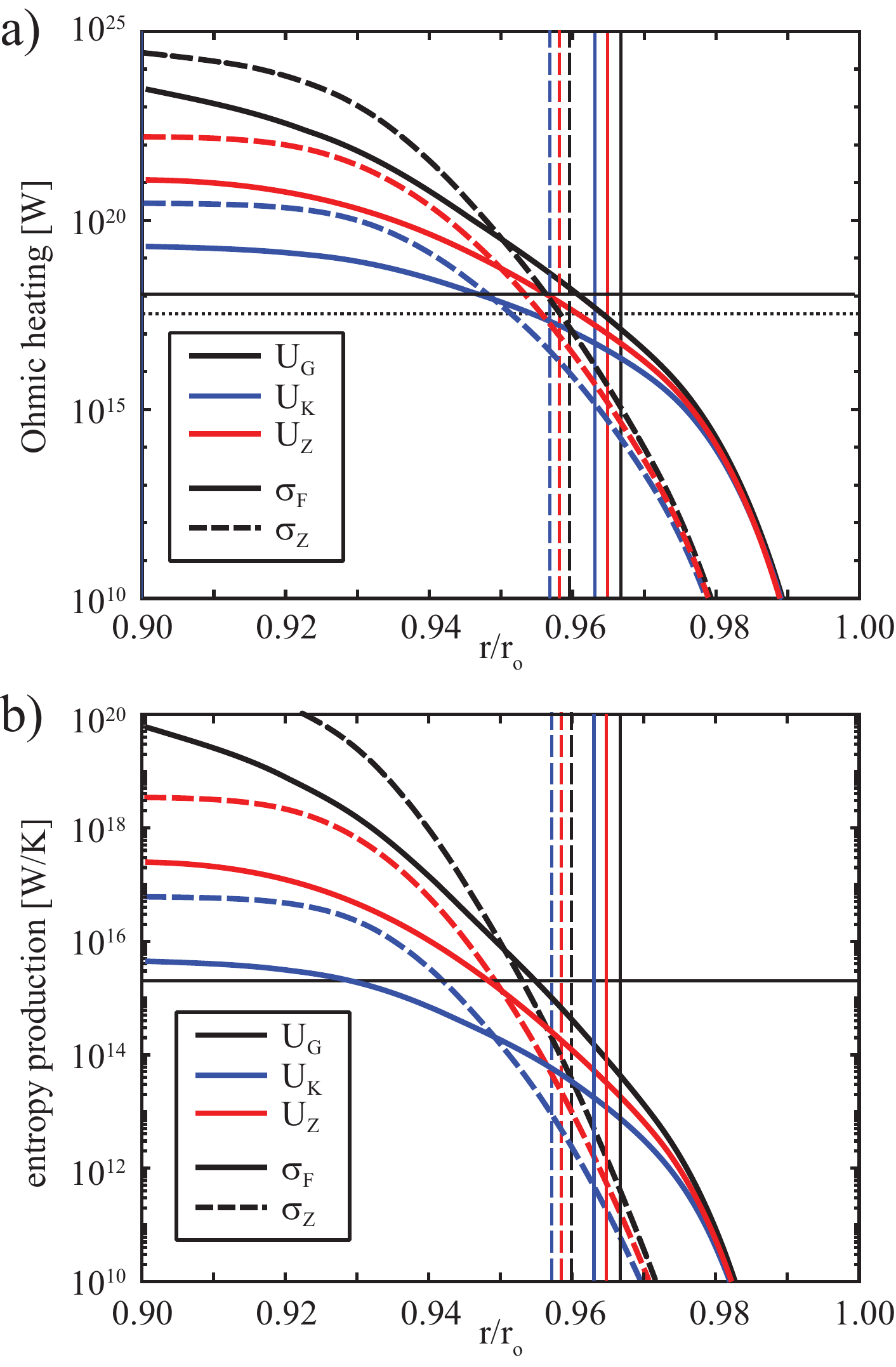}
\caption{Profiles of (a) Ohmic heating   
and (b) entropy production   
in the layer above radius $r$ for current estimate  $\jv^{(O)}$. 
In (a) the solid horizontal line shows 
the total convective power of $1.2\tp{18}\,$W, 
while the dotted horizontal line shows the 
heat flow of $Q_o=3.35\tp{17}\,$W out of Jupiter's interior. 
In (b) the horizontal line indicates the total dissipative entropy 
production predicted by the entropy flux 
$\Ent_T=Q_o/T_o= 2.0\tp{15}\,\mbox{W/K}$ through the outer boundary.
\revi{Vertical lines mark the radii $r_1$ where $\RmL=1$ 
(see \figref{fig:RmL}).} 
}
\label{fig:HeatEnt}
\end{figure} 

Panel a) of \figref{fig:HeatEnt} compares the  Ohmic heating profiles $\Dissi_O(r)$
for the different zonal flow and electrical conductivity models. 
Because of the extremely low conductivity, heating remains negligible in  
outer two percent in radius. 
When using $\jv^{(O)}$, the outermost radius 
where $\Dissi_O$ reaches the level of $\Dissi_T$  is 
$r_\Dissi=0.950\,r_J$ for flow $U_G$ and both conductivity models. 
When using $U_K$ and $\sigma_F$, the Ohmic heating
always remains below $\Dissi_T$. 
Results based on $\jv^{(I)}$ (not shown) are less sensitive to the 
differences between the three flow models at depth and 
are generally similar to the results for $U_G$ and $\jv^{(O)}$. 

The different $r_\Dissi$ values where $\Dissi_O=\Dissi_T$ 
have been  
marked by vertical lines in  \figref{fig:HeatEnt} and  
are listed in column $5$ of \tabref{tab:crossing}.
All are located below the radii $r_1$ where $\RmL=1$ for the respective model combinations (column 3) 
and thus in a region where the approximations employed 
here break down.
The maximum Ohmic heating reached at $r_1$ remains nearly
one order of magnitude below $\Dissi_T$ (see column 8 of  \tabref{tab:crossing}).

Similar inferences hold for the entropy production shown in panel b) of
\figref{fig:HeatEnt}. The entropy condition is less strict 
than the power-based heat condition, and the radii $r_\Ent$  
where the different models exceed the threshold $\Ent_T$ 
 (column 6 of  \tabref{tab:crossing}) are 
somewhat deeper than respective $r_\Dissi$ values.
The largest value of $r_\Ent=0.955$ 
is found for the combination $\U_G$ and $\sigma_Z$. 
The combination of $U_K$ and $\sigma_F$, on the other hand, yields the 
deepest value of $r_\Ent=0.929$. 

The exploration of numerical dynamo simulations by \citet{Wicht2019}
suggest that $\jv^{(O)}$ may provide an acceptable estimate 
for a limited region below $r_1$, at least down to where $\RmL=5$. 
Column 4 of  \tabref{tab:crossing} demonstrates that even the
radius $r_{10}$ where $\RmL=10$ lies deeper than $r_\Dissi$ for 
most flow and conductivity combinations. The only exceptions are the 
results for the geostrophic flow. 
This could indicated that strictly geostrophic flows would indeed violate 
the heating contraint.
\section{Discussion and Conclusion}
\label{sec:Conclusion}

The dominance of Ohmic dissipation in the outer few percent of 
Jupiter's radius leads to simple quasi-stationary dynamo action. 
This can be exploited for estimating the electric currents and the 
Locally Induced Fields with surprisingly high quality \citep{Wicht2019},  
once a conductivity profile, a surface magnetic field model, and flow model are given. 
Here we explored two conductivity profiles, used  
the new Juno-based JRM09 field model, and tested two zonal flow 
models suggested from inversions of Juno gravity measurements.  
A geostrophic zonal flow model was also considered as a third option. 

The estimates roughly apply to the 
upper four percent in radius, or roughly $3000\,$km, where the modified 
magnetic Reynolds number $\RmL$ is smaller 
than one. 
The radial LIF in this quasi-strationary dynamo region 
typically reaches rms values in the order of 
$\mu$T with peak values up to $15\,\mu$T. 
Could such a small contribution be measured by the 
Juno magnetometer? 
The instrument has been designed to provide 
a nominal vector accuracy of $1$ in $10^4$. 
Since the surface field reaches peak values of about 
$2\,m$T, the LIF could indeed be detectable. 

One would still have to separate the LIF 
from contributions produced deeper in the planet. 
What should help with this task, is the 
distinct pattern imprinted by the zonal flows, 
which also leads to a distinct magnetic spectrum.  
The LIF spectrum peaks at degree $\ell=12$ and 
has significant contributions at even higher degrees. 
At $\ell=10$, the largest degree 
provided by JRM09, the LIF amounts to  
about $1$\% of the background field, which seems smaller  
than the estimated JRM09 precision \citep{Connerney2018}. 
Updated future models, based on a larger number of  
Juno orbits, will provide smaller scale details 
and increase the chances of identifying the LIF. 
Another possibility is a dedicated analysis of measurements around 
the 'big blue spot' in JRM09, where 
inductions effects are particularly strong.

Our analysis of Jupiter's heat balance shows 
that Ohmic heating can significantly exceed the 
heat flux $Q_o$ out of the planet's interior. 
Using the interior model by \citet{Nettelmann2012} 
and \citet{French2012} suggests a total 
dissipative heating of $\Dissi_T=3.58\,Q_o = 1.20\tp{18}$W. 

It would be interesting to repeat this 
assessment for the newer Jupiter models that 
include stably stratified regions \citep{Debras2019}.
However, the most important constraint is 
the knowledge of $Q_o$, and the somewhat different 
distribution of internal heat sources implied by the 
newer models can only have a limited effect. 

While the total Ohmic heating remains typically one order
of magnitude below $\Dissi_T$, we find extreme lateral variations. 
Peak values in the Ohmic heating density reach 
$25\,$W/m$^2$ around the 'blue spot' in the JRM09, 
which is nearly five times larger than the mean heat flux 
density from Jupiter's interior. 
These peak values are reached at the bottom 
of the quasi-stationary region, i.e. at a depth 
of $3000\,km$. This is much deeper than any 
(current) remote instrument could reach for. 
For example MWR, the micro-wave instrument on Juno, 
hopes to detect temperature radiation from 
up to $1\,$kbar, which corresponds to a depth 
of about $600\,$km. 
However, the local heating may 
trigger convective plumes that rise to shallower depths 
and thus become detectable. 

We also estimated the entropy flux out of Jupiter's interior
to $2.0\tp{15}\,$W/K. The entropy produced by
zonal-wind related Ohmic 
heating in the quasi-stationary region does not 
exceed this value for any model combination. 
This means that neither Ohmic heating nor the  
entropy production offer any reliable constraint on the 
depth of the zonal winds. 

Below the quasi-stationary region, electric fields become a 
significant contribution to Ohm's law, tend to oppose induction 
effects, and lead to weaker electric currents than predicted 
by our approximations. 
\citet{Wicht2019} demonstrate that the currents remain roughly 
constant below the depth where $\RmL\approx5$ in their numerical 
simulations. However, this may be different in Jupiter where the 
magnetic Reynolds numbers reach values orders of magnitude higher than 
in their computer models.

\Figref{fig:RmL} demonstrates that $\RmL$ increases 
to a value of at least $10^3$ at $0.90\,r_J$. 
This is a consequence of the electrical conductivity profiles 
that easily overcompensate the depth-decrease in zonal flow 
velocities indicated by Juno gravity measurements. 
The zonal flows may thus actually play a larger role for 
dynamo action below than in the quasi-stationary region.
While the gravity data convincingly show that the zonal winds must be significantly weaker below about 
$0.96\,r_J$, they cannot uniquely constrain their structure 
or amplitude at this depth. 

\revi{It has been speculated that the fast observed zonal winds 
may remain confined to a thin weather layer, where differential 
solar heating and also moist convection could 
significantly contribute to the dynamics  
(see for example \citet{Showman2007} or \citet{Thomson2016}). 
\citet{Kong2018} show that the gravity signal can then  
be explained by an independent zonal flow system that 
reaches down to about $0.7\,r_J$ with typical 
amplitudes of about $1\,$m/s and has larger 
latitudinal scales than the surface winds. 
The strongest local dynamo action happens 
towards the bottom of the quasi-stationary region 
where models $U_K$ and $U_Z$ reach velocities of 
about $10\,m/s$. The currents and magnetic fields induced by this  
alternative flow model should thus be roughly 
an order of magnitude weaker than for $U_K$ or $U_Z$. 
Consequently, Ohmic heating and entropy production would be 
two orders of magnitude lower and  play 
practically no role
for the global power or entropy budgets.} 

Below $0.96\,r_J$, full 3d numerical simulations
would be required to model the zonal-wind related
dynamo action. However, since they cannot be run 
at altogether realistic parameters and generally yield 
a much simpler zonal wind pattern, the results 
must be interpreted with care  
\citep{Gastine2014,Jones2014,Duarte2018,Dietrich2018}. 
These simulation suggest that even 
the weaker zonal winds at depth would significantly shear 
the large scale field produced by the 
deeper primary dynamo action. The resulting strong 
longitudinal (toroidal) flux bundles are converted into 
observable radial field by the small scale convective flows
present in this region.  
\revi{The combined action of primary and secondary dynamo 
typically yields a radial surface field that is characterized 
by longitudinal banded structures and  
large scale patches with wavenumber one or two,
resulting in a morphology is often reminiscent of the 
recent Juno-based field model JRM09
\citep{Gastine2014,Duarte2018,Dietrich2018}.}

\bibliography{JWbib}
\bibliographystyle{aa}

\renewcommand{\abstractname}{Acknowledgements}
\begin{abstract}
This work was supported by the German Research Foundation (DFG) in the 
framework of the special priority programs 'PlanetMag' (SPP 1488) 
and 'Exploring the Diversity of Extrasolar Planets' (SPP 1992). 
\end{abstract}

\end{document}